\documentclass[aps,prl,twocolumn,superscriptaddress,showpacs]{revtex4-1}
\usepackage{graphicx}
\usepackage{bm}
\usepackage{color}

\begin{document}
\preprint{APS/123-QED}

\title{Band structure and Fermi surfaces of the reentrant ferromagnetic superconductor Eu(Fe$_{0.86}$Ir$_{0.14}$)$_{2}$As$_{2}$} 

\author{S. Xing}
\email [correspondence : ] {sarah.xing@univ-lorraine.fr}
\affiliation{Institut Jean Lamour, Universit\'{e} de Lorraine, UMR 7198 CNRS, BP70239, 54506 Vandoeuvre l\`{e}s Nancy, France}
\author{J. Mansart}
\affiliation{Laboratoire de Physique des Solides, Universit\'{e} de Paris-Sud, Paris, France}
\author{V. Brouet}
\affiliation{Laboratoire de Physique des Solides, Universit\'{e} de Paris-Sud, Paris, France}
\author{M. Sicot}
\affiliation{Institut Jean Lamour, Universit\'{e} de Lorraine, UMR 7198 CNRS, BP70239, 54506 Vandoeuvre l\`{e}s Nancy, France}
\author{Y. Fagot-Revurat}
\affiliation {Institut Jean Lamour, Universit\'{e} de Lorraine, UMR 7198 CNRS, BP70239, 54506 Vandoeuvre l\`{e}s Nancy, France}
\author{B. Kierren}
\affiliation {Institut Jean Lamour, Universit\'{e} de Lorraine, UMR 7198 CNRS, BP70239, 54506 Vandoeuvre l\`{e}s Nancy, France}
\author{P. Le F\`{e}vre}
\affiliation{Synchrotron SOLEIL, UR1 CNRS, Saint-Aubin-BP48, 91192 Gif-sur-Yvette, France}
\author{F. Bertran}
\affiliation{Synchrotron SOLEIL, UR1 CNRS, Saint-Aubin-BP48, 91192 Gif-sur-Yvette, France}
\author{J. E. Rault}
\affiliation{Synchrotron SOLEIL, UR1 CNRS, Saint-Aubin-BP48, 91192 Gif-sur-Yvette, France}
\author{U. B. Paramanik}
\affiliation{Department of Physics, Indian Institute of Technology, Kanpur 208016, India}
\author{Z. Hossain}
\affiliation{Department of Physics, Indian Institute of Technology, Kanpur 208016, India}
\author{A. Chainani}
\affiliation{Institut Jean Lamour, Universit\'{e} de Lorraine, UMR 7198 CNRS, BP70239, 54506 Vandoeuvre l\`{e}s Nancy, France}
\affiliation{RIKEN SPring-8 Center, 1-1-1 Sayo-cho, Hyogo 679-5148 Japan}
\affiliation{National Synchrotron Radiation Research Center, Hsinchu 30076,  Taiwan}
\author{D. Malterre}
\affiliation{Institut Jean Lamour, Universit\'{e} de Lorraine, UMR 7198 CNRS, BP70239, 54506 Vandoeuvre l\`{e}s Nancy, France}
\date{\today} 

\begin{abstract}
The electronic structure of the reentrant superconductor Eu(Fe$_{0.86}$Ir$_{0.14}$)$_{2}$As$_{2}$ (T$_c$ = 22 K) with coexisting ferromagnetic order (T$_M$ = 18 K) is investigated using angle-resolved photoemission spectroscopy (ARPES) and scanning tunneling spectroscopy (STS). We study the in-plane and out-of-plane band dispersions and Fermi surface (FS) of Eu(Fe$_{0.86}$Ir$_{0.14}$)$_{2}$As$_{2}$. The near E$_F$ Fe 3d-derived band dispersions near the $\Gamma$ and X high-symmetry points show changes due to Ir substitution, but the FS topology is preserved. From momentum dependent measurements of the superconducting gap measured at T = 5 K, we estimate an essentially isotropic s-wave gap ($\Delta\sim5.25\pm 0.25$ meV), indicative of strong-coupling superconductivity with 2$\Delta$/k$_{B}$T$_{c}\simeq$ 5.8.
The gap gets closed at temperatures T $\geq$ 10 K, and this is attributed to the resistive phase which sets in at T$_M$ = 18 K due to the  Eu$^{2+}$-derived magnetic order.  The modifications of the FS with Ir substitution clearly indicates an effective hole doping with respect to the parent compound.

\end{abstract}


\maketitle

\section{Introduction}

Since the discovery of superconductivity in the iron-based pnictides \cite{Kamihara}, extensive experimental and theoretical studies have revealed their unusual properties as well as their similarities and differences with the cuprate superconductors \cite{Rotter,Cruz,Ding,Borisenko,Brouet,Mazin,Kuroki,Kruger, Shimojima,Yi,Maiti}. While the parent compounds of the so-called 122 series AFe$_2$As$_2$ (A = Ba, Ca, Sr, Eu etc.) are metals \cite{Rotter,Cruz}, they exhibit a spin density wave (SDW) transition at temperatures ranging from T$_{SDW}\sim140 K - 190 K$. This is in contrast to the undoped cuprates such as La$_2$CuO$_4$ and Nd$_2$CuO$_4$, which are antiferromagnetic (AF) insulators, specifically classified as charge-transfer insulators \cite{Fujimori,Shen}. Further, the FS of the iron pnictides are dominated by d$_{xy}$, d$_{yz}$ and d$_{zx}$ character bands \cite{Ding,Borisenko,Brouet}, although a small d$_{z^2}$ contribution was also found at the ($\pi,\pi$)-point in the Brillouin zone\cite{Hashimoto,Kasahara}.

For the iron-based 122 systems, A-site hole-doping can stabilize superconductivity up to T$_c\sim 40 K$. In addition, they also show superconductivity for substitutions (Co, Ni or Cu) on the Fe-site. Interestingly, ARPES studies showed that Co substitution in BaFe$_2$As$_2$ (Ba122) leads to electron-doping proportional to the substitution content x \cite{Brouet2009,Brouet2012,Liu,Malaeb,Thirupathaiah} whereas for Ni and Cu substitution, it was found that the number of electrons that participate in the FS formation gets reduced and this was attributed to the increasing impurity potential of the substituent \cite{Ideta}.
Furthermore, the iron-based systems usually exhibit a structural transition from a tetragonal to orthorhombic structure at a temperature T$_s$ in the vicinity of, or even coinciding with, T$_{SDW}$. This transition is accompanied by significant changes in the FS \cite{Liu,Zhang, Shimojima14}. In particular, the splitting of the saddle band and the FS droplet structure at the X($\pi,\pi$) point vanish above T$_s$ = T$_{SDW}$ = 138 K in Ba122 \cite{Yi09}. 

The substitution of Fe by Ir (Ir is isovalent to Co) in EuFe$_2$As$_2$ (Eu122) results in an intriguing reentrant ferromagnetic (FM) superconducting (SC) phase \cite{Paramanik}.  Eu122 is a paramagnetic metal at room temperature forming in the tetragonal structure. It undergoes a SDW transition at T$_{SDW}\! =\!190$ K, which coincides with a transition to an orthorhombic structure. Eu122 also shows an AF transition at T$_N\!=\!$ 18 K due to the Eu$^{2+}$ magnetic moments \cite{Paramanik}. 
ARPES studies on Eu122 showed that the FS exhibits clear modifications, similar to Ba122, with the X($\pi,\pi$)-centered electron FS getting transformed into smaller FS droplets below T$_s$ = T$_{SDW}$ = 190 K \cite{Jong,Zhou}.
While the high temperature SDW transition as well as the tetragonal to orthorhombic structural transition get suppressed with Ir substitution, leading to the onset of superconductivity, the Eu$^{2+}$ derived AF order gets transformed to FM ordering below T$_M$ $\sim$18 K in Eu(Fe$_{0.86}$Ir$_{0.14}$)$_{2}$As$_{2}$ (Ir-Eu122) \cite{Paramanik,Anand}.
Most interestingly, a reentrant superconductivity is observed: Ir-Eu122 exhibits a SC transition at T$_c$ = 22 K, followed by a FM order at T$_M\!=\!$ 18 K. From 18 K to about 10 K, the superconductivity gets suppressed and a resistive phase is observed, but below $\sim$10 K, the system re-enters the SC state as confirmed by a diamagnetic signal and resistivity measurements \cite{Paramanik}. Neutron diffraction and $\mu SR$ studies on Ir-Eu122 have shown that the 
 FM Eu moments are aligned along the c-axis down to 1.8 K, thus indicating its coexistence with the reentrant superconductivity. The absence of both, the Fe moment ordering and the structural transition was also confirmed \cite{Anand}. 
 
 \begin{figure}[!t]
\includegraphics[width=8cm, keepaspectratio]{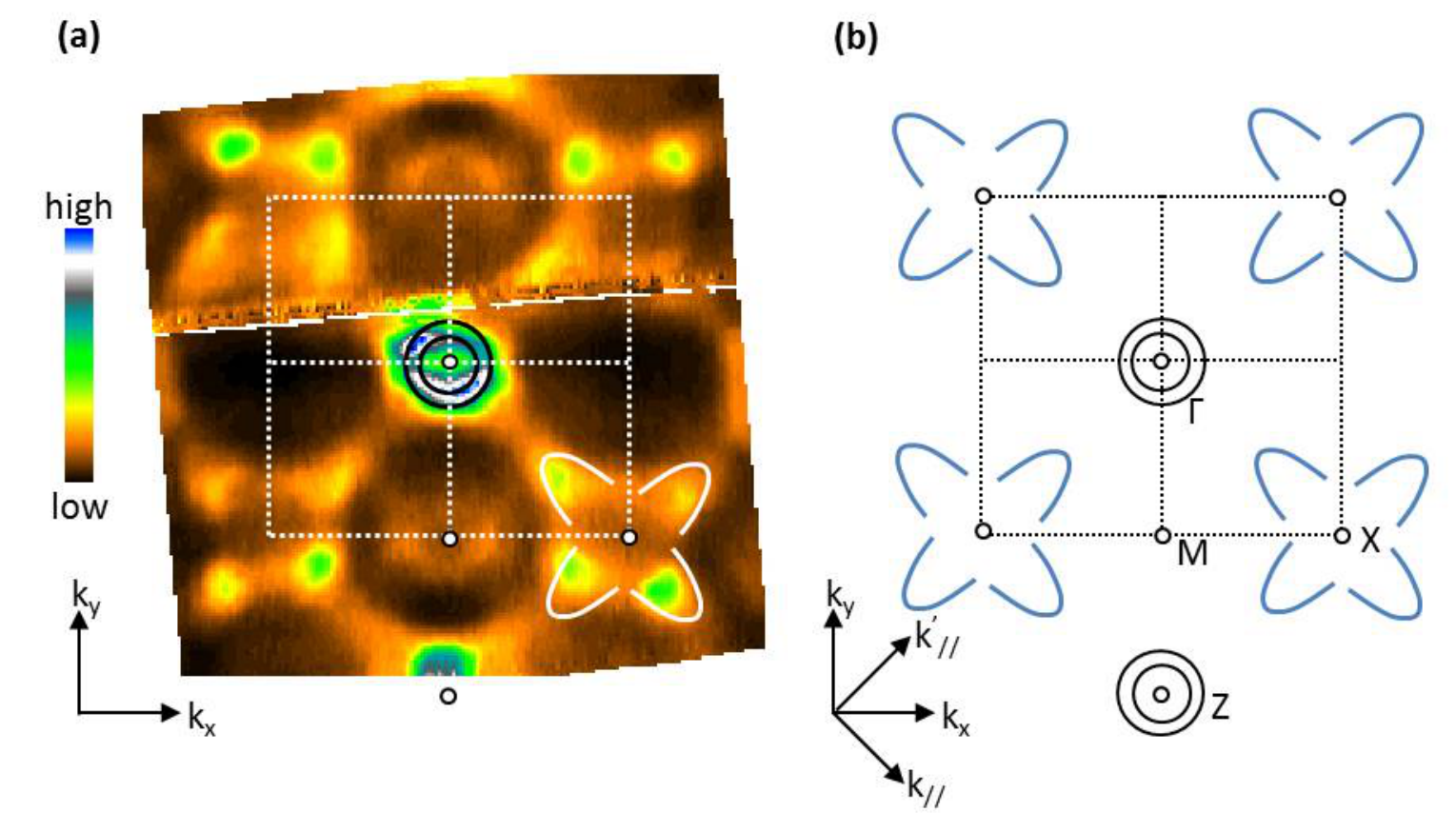}
\caption{(Color online) (a) Fermi surface measured at 5 K using 100 eV photon energy and LV polarization. (b) Schematic shape of the FSs with hole pockets at $\Gamma$ point and droplet-like features around X point. 
}\label{Fig.one}
\end{figure} 

We have carried out ARPES measurements to determine the in-plane and out-of-plane band dispersions and FSs of Ir-Eu122.  We also investigate the SC gap behavior using ARPES and STS measurements. The results show that the SC gap seen at T $\!=\!$ 5 K gets closed at temperatures T $\geq$ 10 K, thereby confirming the resistivity measurement and re-entrant superconductivity  \cite{Paramanik}.
We also show that in contrast to naive expectations, the Ir substitution leads to hole doping behavior. Both the $\Gamma$ and X pockets are shifted in energy with respect to the parent compound.  These modifications of the FS can be associated with the suppression of the SDW and structural transition in Ir-Eu122, and reveal the interplay between magnetic ordering and band structure due to Ir substitution. 
       
\section{Sample preparation, characterization and experimental details}

Eu122 and Ir-Eu122 crystallize in the tetragonal I4/mmm structure. The lattice parameters are a=3.9113 $\AA$ and c=12.136 $\AA$ for the parent compound Eu122 and a=3.9365 $\AA$ and c=12.027 $\AA$ for Ir-Eu122 \cite{Paramanik,Anand}.Single crystals of Ir-Eu122 were grown from self-flux using conventional high temperature solution growth techniques, as described in reference \cite{Qi}. The FeAs and IrAs precursors were synthesized by the reaction of Fe powder or Ir powder and As chips at 500 $^{\circ}$C for 10 h and then 900 $^{\circ}$C for 20 h in a sealed tantalum tube. The precursors and Eu were placed in an alumina crucible, and sealed inside a tantalum tube. The assembly was heated to 1200 $^{\circ}$C slowly and held there for 5 h, and then was cooled to 1030 $^{\circ}$C at a rate of 3 $^{\circ}$C h$^{-1}$; finally it was furnace-cooled to room temperature. Several plate-like single crystals of dimension $\sim$4$\times$4$\times$0.15 mm$^{3}$ were obtained. The samples were characterized by X-ray diffraction to determine the phase purity and crystal structure. A scanning electron microscope (SEM) equipped with energy dispersive X-ray (EDX) analysis was used to examine the homogeneity and composition of the samples. Resistivity, ac magnetic susceptibility measurements and heat capacity measurements were carried out to characterized the physical properties. ARPES experiments were carried out at the CASSIOPEE beamline at the SOLEIL synchrotron with a Scienta R4000 analyzer. The angular resolution was 0.3 degree and the energy resolution better than 15 meV. The temperature dependence of the superconducting gap was recorded with the ultimate resolution of the beamline (8 meV). The linearly polarized light was used to select even and odd parity bands with respect to the emission plane. STS measurements were carried out at 5 K with a lock-in detection in open feedback loop conditions at a frequency of 700 Hz and with a bias modulation of 2 mV \textit{rms}. The shiny plate-like Ir-Eu122 samples were cleaved just before spectroscopic (ARPES and STS) measurements. Measurements of the superconducting gap at T = 5 K indicates essentially isotropic s-wave gap ($\Delta\sim5.25\pm 0.25$ meV), corresponding to strong-coupling superconductivity (2$\Delta$/k$_{B}$T$_{c}\simeq$ 5.8).

\section{Results and discussion}

\begin{figure}[h!]
\begin{center}
\includegraphics[scale=0.40]{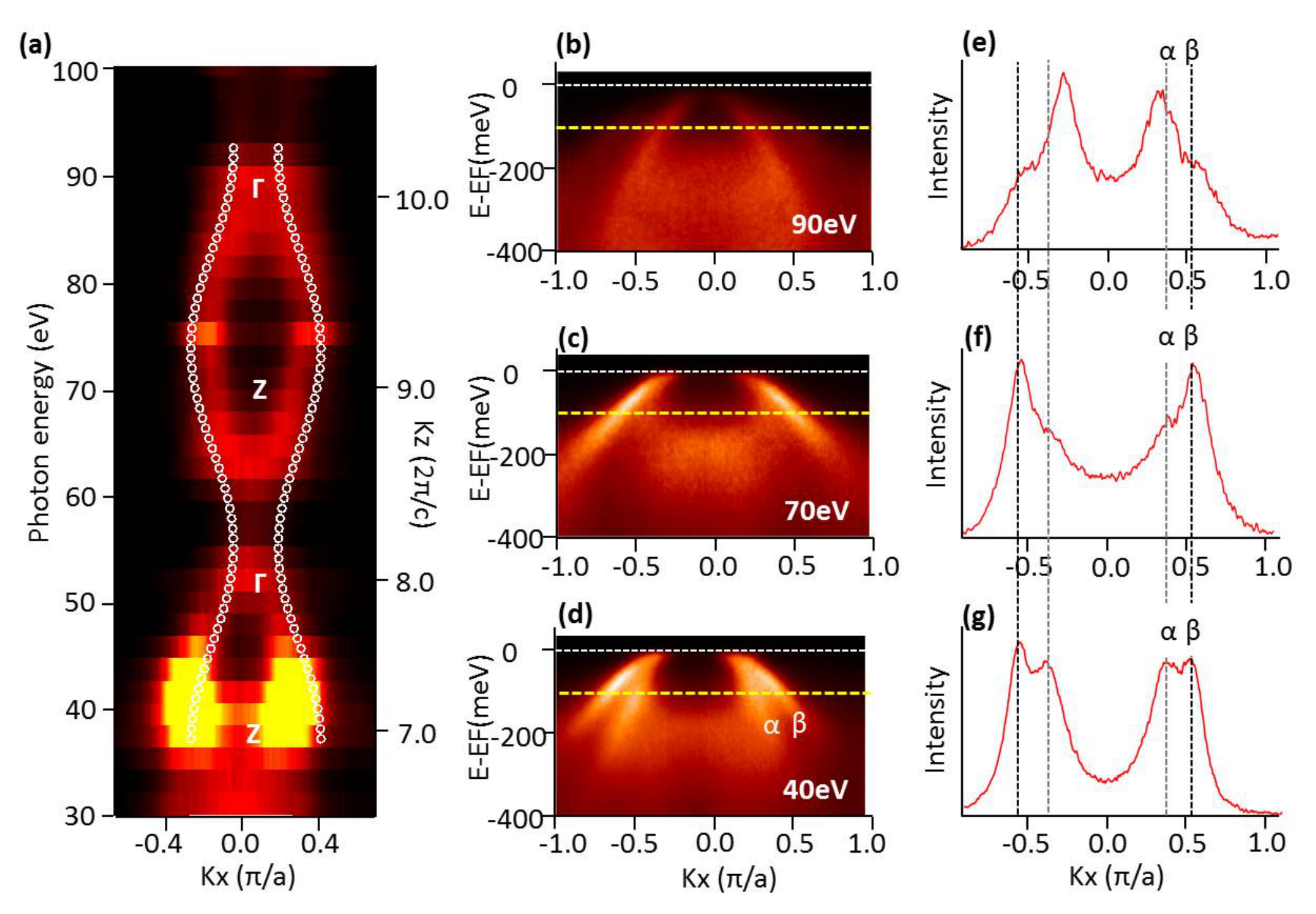}
\end{center}
\caption{(Color online) ARPES measurement of Eu(Fe$_{0.86}$Ir$_{0.14}$)$_{2}$As$_{2}$2 carried at 5 K, with LH polarization. (a) Photoemission spectral intensity at E$_{F}$ measured along k$_{x}$ as a function of photon energy or, equivalently, of k$_{z}$. Open symbols in (a) are a guide to the eye. (b-d) (E-k) maps taken at 90 eV, 70eV and 40 eV. Yellow dashed lines are drawn at 100 meV binding energy. Corresponding MDCs are depicted in (e-g).}
\end{figure}

The electronic structure and FS of iron pnictides are characterized by the contribution of several bands close to the Fermi energy. 
The momentum map at $E_F\pm$ 10 meV recorded with 100 eV photon energy (with linear vertical (LV) polarization) is presented in Fig. 1 (a). It shows that the FS is composed of two concentric hole pockets around the $\Gamma$ and Z points, and two elongated pockets at X point as schematically illustrated in Fig. 1 (b), similar to the FS of Eu122 \cite{Jong,Zhou}.
\begin{figure*}[t]
  \includegraphics[width=\textwidth]{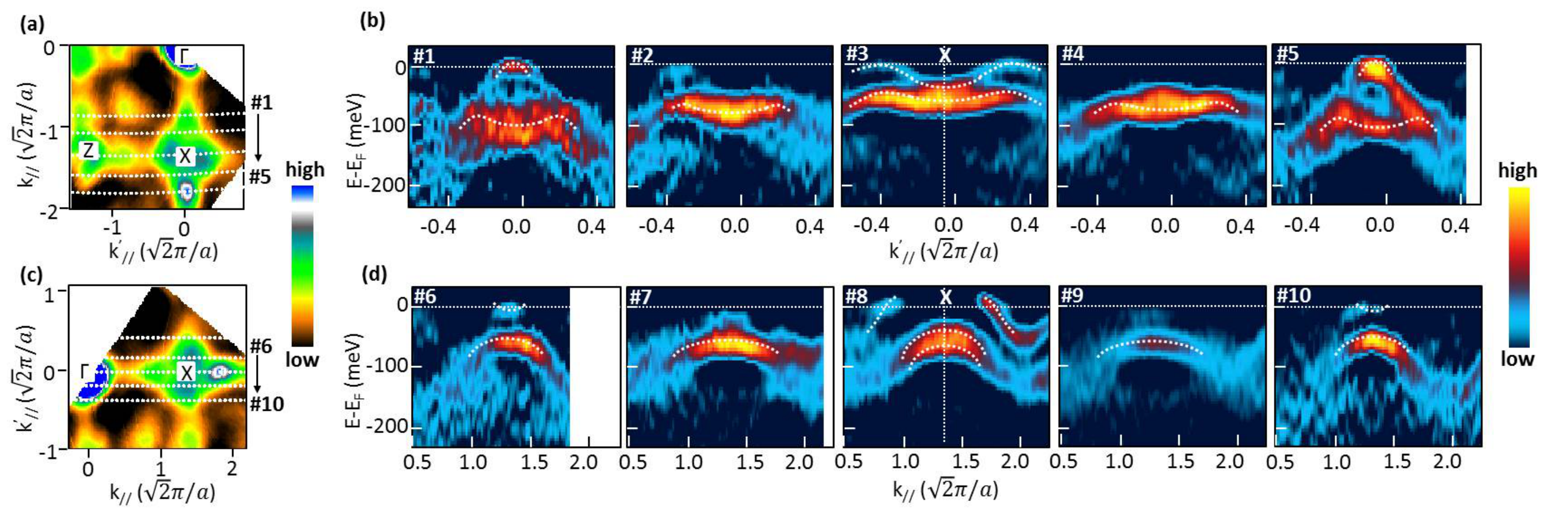}
  \caption{(Color online) Electronic structure around X point at 5 K. Measurements are carried out at 83 eV photon energy with LV polarization.  (a) Fermi surface map, cuts 1-5 are performed parallel to Z-X direction as indicated. (b) The second derivative (E-k) maps along cuts 1-5. (c) Fermi surface map and cuts 6-10 parallel to $\Gamma$-X direction. (d) Corresponding second derivative (E-k) maps.}
\end{figure*}
Fe-based superconductors are usually characterized by a very anisotropic electronic structure with strong two-dimensional dispersions in the Fe  planes but a small perpendicular dispersion can also be observed. 
 In order to quantify the 3-dimensional character of the electronic states in Ir-Eu122 close to the Fermi energy ($E_F$), we investigated the k$_z$ dependence of the band states from ARPES measurements at T=5 K, in the SC phase. In Fig. 2 (a), we show the momentum distribution curves (MDC) at $E_F$  in the k$_x$ direction for k$_y$=0 and for photon energies between 30 and 100 eV. In Fig. 2 (b,c,d)  we present the (E-k) maps collected along $\Gamma$-M direction for 90 eV, 70 eV and 40 eV photon energies. These maps clearly show two bands with hole character, labeled $\alpha$ (inner) and $\beta$ (outer) bands.
The more intense structure in Fig. 2 (a) corresponding to $\alpha$ band exhibits a clear $k_z$ dependence indicating its 3-dimensional character. The open symbols represent a fit of the spectral intensity maxima by a sinusoidal function of $k_z$ and reflects the dispersion in the $\Gamma$-Z direction.
Indeed, at normal emission, k$_z$ is simply related to the photon energy, the work function $W$ and the inner potential V$_{0}$. Using an inner potential $V_0$=15 eV, we could obtain a good match to the observed periodic behavior as a function of the $2\pi/c$ reciprocal vector, in good agreement with other iron pnictides \cite{Brouet,Brouet2009}. We could thus obtain the wave vectors corresponding to the Z points ( k$_{F}$ maxima) and the $\Gamma$ points ( k$_{F}$ minima). The $\beta$ band presents a very small $k_z$ dependence. This is corroborated by the MDCs of Fig. 2 (e-g) recorded at 100 meV B.E. corresponding to dashed yellow lines on the maps  in Fig.2 (b-d). They clearly evidence the inner $\alpha$ and outer $\beta$ bands for the three photon energies but with different spectral weight and different k$_{z}$-dispersion. A very small dispersion is observed for the outer $\beta$ band in contrast to the larger dispersion of $\alpha$ band, as is usually observed in Fe-pnictides \cite{Brouet2009}. This behavior results from the orbital character of these hole bands. As deduced from polarization dependence of ARPES intensity in an earlier study \cite{Brouet2012}, the inner structure is associated with two nearly degenerate bands with d$_{xz}$ and d$_{yz}$ character, whereas the outer band corresponds to d$_{xy}$ orbitals characterized by a smaller hybridization in the $z$ direction. The very large difference in spectral weight observed at the same point (Z point in Fig. 2 (b,d)) was already observed in the parent compound \cite{Richard} and is likely due to matrix element effects. 

Comparison with EuFe$_{2}$As$_{2}$  \cite{Zhou,Adhikary} shows an increase of the Fermi momenta in units of $(\pi/a)\AA^{-1}$ for the $\alpha$ (0.12 $\rightarrow$ 0.2) and $\beta$ (0.25 $\rightarrow$ 0.29) bands with Ir substitution.
This increase of k$_{F}$ for $\alpha$ and $\beta$ bands could be naively interpreted  as hole-doping whereas the electronic structure is expected to be electron-doped when Fe is replaced with Ir (one electron more). Moreover a change of the Fermi velocities indicates a decrease of correlation effects as was also observed in Ru-doped 122 \cite{Brouet, Xu12}. These behaviors suggest that the evolution of the electronic structure does not follow the rigid band picture. This is confirmed by supercell band-structure calculations for Eu(Fe$_{0.875}$Ir$_{0.125}$)$_{2}$As$_{2}$ and compared with EuFe$_{2}$As$_{2}$, as reported earlier \cite{Paramanik}. Moreover, in a systematic study of the modifications of the electron density associated with various substituents (Co, Ni, Cu, Zn, Ru, Rh and Pd) in BaFe$_{2}$As$_{2}$ and FeSe, Wadati et al. \cite{Wadati} found by supercell band-structure calculations that  the additional charges due to substitution remain localized in the Muffin-Tin sphere at the substituted sites. Further, our experiments show that 
in addition to the hole-pockets around the $\Gamma$ point, bands in the vicinity of the X point show complex FS topology. Second derivative band-maps in $\Gamma$-X and Z-X directions, and parallel to these directions are presented in Fig. 3.
At X point, two bands are observed with inverted curvatures along perpendicular directions (cuts 3 and 8). i.e. a saddle point behavior. The character of the saddle point observed at X point is similar to that reported in the parent compounds Eu122 and Ba122 \cite{Zhou,Yi2009}. Nevertheless, the two saddle bands at X are shifted by 30-40 meV towards the Fermi energy with respect to Eu122. This corroborates the hole-doping induced by Ir substitution already observed on the bands at $\Gamma$. Moreover, the existence of a splitting in two bands, which has been shown to be related to the structural and magnetic transition \cite{Yi2009}, is surprising in the non-magnetic tetragonal Ir-Eu122 material.
Cuts parallel to Z-X direction are depicted in Fig. 3(b). Moving away from X point (cut 2 and cut 4), the shallow electron band disappears whereas the deeper one remains at the same energy position. Going further, a tiny hole band emerges, and gives rise to the droplet like FS. A similar behavior is observed for the cuts parallel to the $\Gamma$-X direction (Fig. 3(c,d)) with the inversion of the shallow hole band at X into an electron-like droplet. This droplet is usually found in the Eu122 orthorhombic phase and disappears above T$_s$. But the observed FS topology has been discussed for highly hole-doped K$_x$Ba$_{1-x}$Fe$_2$As$_2$ (x$>$ 0.85) system \cite{Khan}.

\begin{figure}[t]
\begin{center}
\includegraphics[scale=0.33]{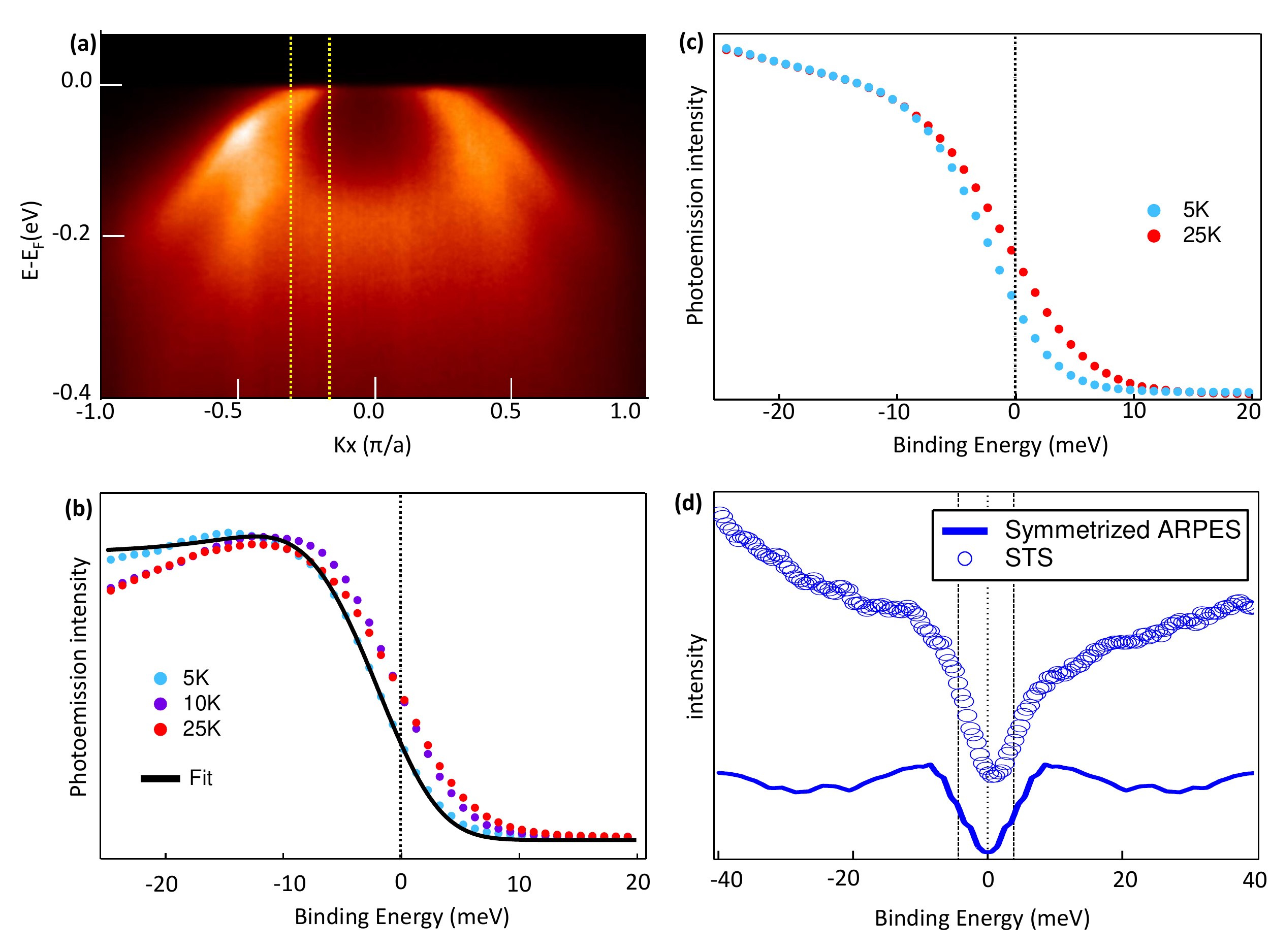}
\end{center}
\caption{(Color online) (a) Band dispersion at 5 K of Eu(Fe$_{0.86}$Ir$_{0.14}$)$_{2}$As$_{2}$ in $\Gamma$-M direction taken at 39 eV photon energy, LH polarization. (b) Integrated photoemission intensities over $k_{x}\!=\!-0.30$ and $k_{x}\!=\!-0.15$ in $\pi/a$ unit (yellow lines in a) at 5 K, 10 K and 25 K. The leading edge is clearly shifted when the temperature changes from 10 K to 5 K. The gap magnitude  ($\Delta$=5.5 meV) is obtained by using a Dynes' function fitting (solid line).  (c) Integrated photoemission intensities over $k_{x}\!=\!-1$ and $k_{x}\!=\!1$ in units of $\pi/a$ for 5 K and 25 K, showing the SC gap. (d) The T = 5 K symmetrized ARPES (solid line) and the T = 5 K STS  (open circles) spectra show consistent gap values.} 
\end{figure}
In order to evidence the gap opening and its evolution with temperature, we measured the ARPES spectra corresponding to the two $\Gamma$-centered hole bands with the ultimate energy resolution (8 meV) of the CASSIOPEE beamline at 5 K (the reentrant region), 10 K (the magnetism-induced resistive region ; we have also measured the spectrum at 15 K but it is very similar to the 10 K spectrum and is not shown here)  and 25 K (the normal region above T$_c$). The energy has been carefully calibrated by measuring the Fermi step of a Au foil just before and after each measurement. The 5 K map is shown in Fig. 4 (a). In Fig. 4 (b), we plot the spectral intensity integrated in a narrow momentum range $\Delta k_x$=0.15 $(\pi/a)\AA^{-1}$ around the Fermi momentum (between the yellow dotted lines in Fig. 4 (a)). The 25 K-spectrum is characterized by a Fermi step broadened by the Fermi function and the energy resolution. At 10 K, the leading edge is narrower (Fermi function effect) but appears at the same energy. However, at 5 K, the leading edge is shifted to lower energy by about 2 meV revealing the opening of the SC gap in this momentum range.

Fig. 4(c) presents the photoemission intensity integrated along k$_{x}$ ($-1\!\le\! k_{x}\!\le\!1$), and the spectrum exhibits exactly the same shift of the leading edge at T = 5 K. The symmetrized ARPES spectrum at T = 5 K corresponding to a narrow momentum range $\Delta k_x=$ 0.15 $(\pi/a)\AA^{-1}$ around the Fermi momentum is also plotted in Fig. 4 (d) and compared with the STS spectrum recorded at T = 5 K. Both spectra are very similar and as STS integrates over a broad momentum range, the comparison suggests that the gap opens for all momenta of the hole pockets near $\Gamma$ as expected for a s-wave gap. In order to confirm the same, and as it is well known that the leading edge shift underestimates the gap \cite{Ding, Zhang-b}, the gap magnitude was estimated by a fitting procedure using the Dynes' function for an s-wave gap \cite{Dynes}. In this approach, the spectral density depends on two parameters $N(E)=Re\Big((E-i\Gamma)/[(E-i\Gamma)^2-\Delta^2]^{1/2}\Big)$, where $\Delta$ is the gap parameter and $\Gamma$ is a broadening term due to finite lifetime of the excitations. The solid line in Fig. 4(b) represents the Dynes' function obtained with $\Delta$ = $5.25\pm  0.25$ meV and $\Gamma$ = 3 meV. A very good agreement with the 5 K-experimental spectrum is obtained, and the gap value (2$\Delta$/k$_{B}$T$_{c}\simeq$ 5.8) indicates that the gap is beyond the BCS weak-coupling limit and Ir-Eu122 is better described as a strong coupling superconductor.

In addition, it is necessary to check the momentum dependence of the gap to determine the gap symmetry.  This is particularly important as diverse types of gaps, nodal and nodeless (including isotropic and anisotropic variants) have been observed in iron pnictide based superconductors \cite{Ding, Hashimoto, Zhang-b}.

In order to study this gap character in Ir-Eu122, we present in Fig. 5 photoemission spectra integrated over the same momentum range as discussed previously ($\Delta k$=0.15 $(\pi/a)\AA^{-1}$ around k$_F$) for different polar angles ($0 \le \theta \le$ 45$^{\circ}$). By symmetry, analysis with polar angles varying between 0$^{\circ}$ and 45$^{\circ}$ is enough for investigating the momentum dependence of the superconducting gap. 
These spectra were recorded with a lower energy resolution than in Fig. 4 ($\Delta E=$  14 meV) but sufficient to evidence the gap opening.
On the FS around $\Gamma$ point, all spectra are superimposed leading to the conclusion that the gap is nodeless and isotropic. The leading edges are found at -2 meV in good agreement with the high resolution measurement of Fig. 4 and the spectra can be simulated by a Dynes' function with the same $\Delta$ and $\Gamma$ parameters. (insert of Fig. 5). The observed isotropic gap is also consistent with the empirical relation \cite{Hashimoto, Xia} of a nodeless gap when the pnictogen height h$_{Pn}$ at room temperature, is $>$ 1.33  $\AA$, since h$_{Pn}$ = 1.34  $\AA$ for Eu(Fe$_{0.86}$Ir$_{0.14}$)$_{2}$As$_{2}$.

 Thus, although Ir-Eu122 undergoes a SC transition at T$_c$=22 K,  the FM ordering of Eu$^{2+}$ moments competes with superconductivity and causes a resistive behavior between 18 K and ~10 K \cite{Paramanik} which prevents superconductivity and the formation of a SC gap. Below 10 K, the pairing interactions are presumably stronger than magnetic interactions and result in reentrant superconductivity coexisting with FM order, with the observation of a completely opened gap.
A similar behavior of the T-dependence of the gap was very recently reported for the rentrant superconductor Eu(Fe$_{0.93}$Rh$_{0.07}$)$_{2}$As$_{2}$ using optical spectroscopy \cite{Baumgartner}.  

\begin{figure}[t]
\begin{center}
\includegraphics[width=8cm, keepaspectratio]{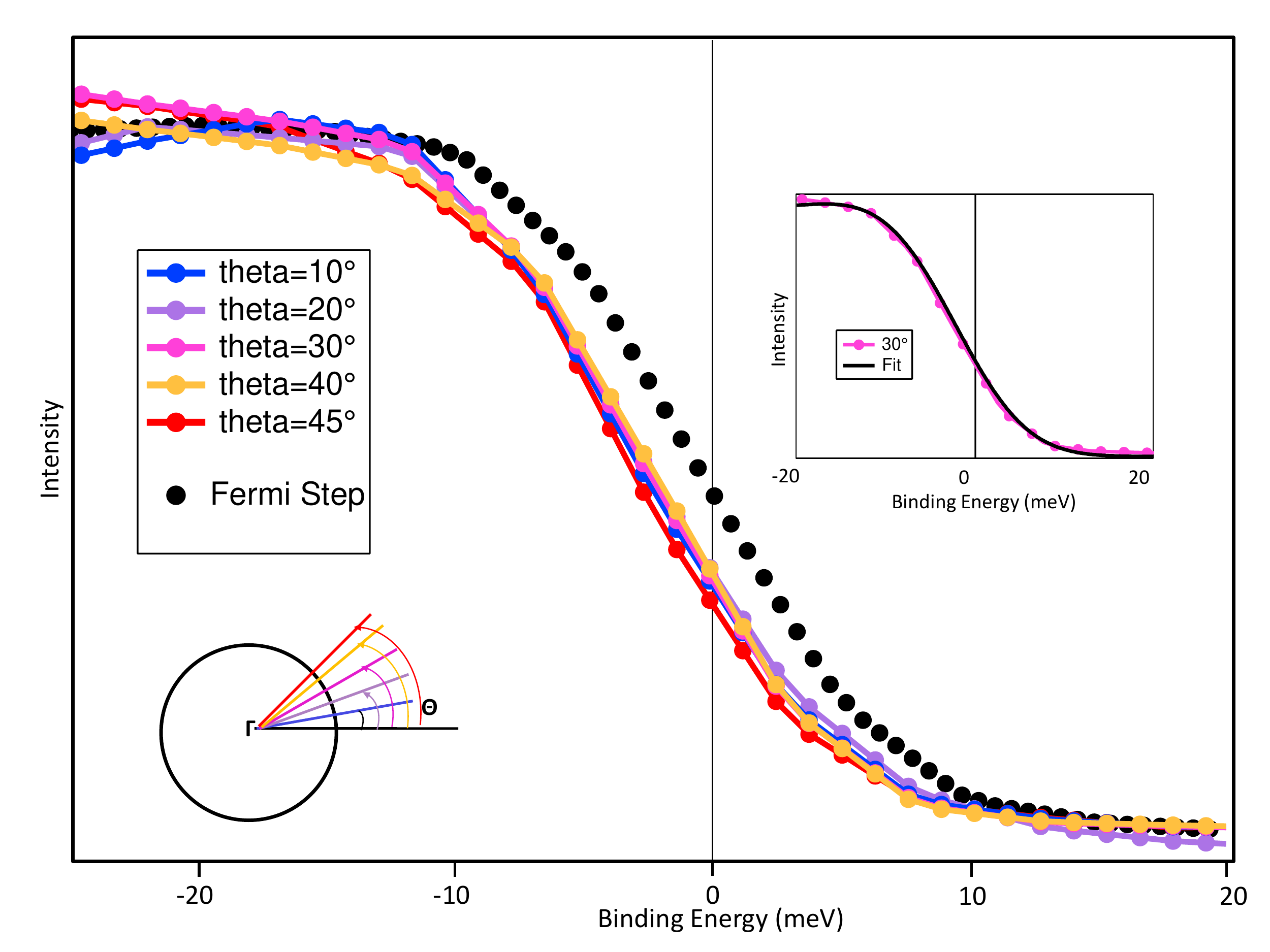}
\end{center}
\caption{(Color online) Integrated photoemission intensities over $\Delta k\!=\!-0.15$ in $\pi/a$ unit for different polar angles $\theta$. The different spectra are superimposed with a leading edge at 2meV below E$_F$. The dashed line is a Fermi edge for a metal surface at the same temperature. Insert : 30$^{\circ}$ experimental spectrum (dotted line) and a Dynes' function with $\Delta=5.25\pm  0.25$ meV and $\Gamma=$ 3 meV (solid line). } 
\end{figure}
\section{Conclusion}

In this study, we have investigated the electronic structure of optimally doped reentrant FM superconductor Eu(Fe$_{0.86}$Ir$_{0.14}$)$_{2}$As$_{2}$. We clarify the subtle interplay between superconductivity and magnetism. We evidence an isotropic gap at 5 K whose magnitude is in agreement with the BCS relation with T$_c$ = 22 K. However, this gap is suppressed by Eu FM order at T $\geq$ 10 K.
The results reveal a new puzzle in an Fe-pnictide : although Ir substitution suppresses both the Fe moment derived SDW transition and the tetragonal to orthorhombic structural transition, the SC state develops with a FS topology which is similar to the one observed in the orthorhombic magnetic phase of Eu122. In particular, the splitting of the saddle bands at the X point which is usually considered as the signature of the orthorhombic structure, is surprisingly present in the tetragonal non-magnetic Ir-Eu122. Moreover, the evolution of the hole-bands close to $\Gamma$ and electron and hole pockets close to X clearly indicates that the change of band positions and Fermi momenta with respect to the parent compound corresponds to a removal of states close to $E_F$ whereas an electron doping is expected from the relative electron configuration of Fe and Ir. This behavior shows that the evolution of the band structure cannot be simply described as doping in a rigid band picture, as was also found in supercell DFT calculations \cite{Paramanik,Wadati}.
This apparent hole doping of Ir-Eu122 is corroborated by the similarities with the FS of hole-doped K$_x$Ba$_{1-x}$Fe$_2$As$_2$ (x$>$ 0.85) systems which exhibit non-magnetic tetragonal structure with a propeller-shaped FS at the X point \cite{Khan}.
Our results shed new light on the electronic properties of 122 Fe-pnictides. Ir-Eu122 exhibits the characteristic features of the parent compound FS in spite of the absence of the $C_4$ symmetry breaking associated with the tetragonal orthorhombic transition. Moreover, the shift of the bands at the $\Gamma$ and X points strongly suggests a hole doping and then the distribution in energy and momentum of the additionally expected electron due to the Ir substitution remains an open question. \\

Acknowledgements : A.-C. thanks the Institut Jean Lamour, Universit\'{e} de Lorraine for hospitality and financial support during the course of this work. A. -C. also thanks the Ministry of Science and Technology of the Republic of China, Taiwan for financially supporting this research under contract No. MOST 106-2112-M-213-001-MY2. \\

\end{document}